# Diffusion-weighted SPECIAL improves the detection of J-coupled metabolites at ultra-high magnetic field

J. Mosso[1,2,3], D. Simicic[2,3], B. Lanz[2,3], R. Gruetter[1], C. Cudalbu[2,3]

[1] LIFMET, EPFL, Lausanne, Switzerland
[2] CIBM Center for Biomedical Imaging, Switzerland
[3] Animal Imaging and Technology, EPFL, Lausanne, Switzerland

Corresponding author: Jessie Mosso (jessie.mosso@epfl.ch)

Word count: 4997 excluding the Acknowledgements

# Abstract

## Purpose

To improve the detection and subsequent estimation of the diffusion properties of strongly J-coupled metabolites in diffusion-weighted magnetic resonance spectroscopy (MRS).

## Methods

A new sequence for single-voxel diffusion-weighted $^1$H MRS (DWS), named DW-SPECIAL, is proposed. It combines the semi-adiabatic SPECIAL sequence with a stimulated echo (STE) diffusion block. Acquisitions with DW-SPECIAL and STE-LASER, the current gold-standard for rodent DWS experiments at high fields, were performed at 14.1T on phantoms and in vivo on the rat brain. The apparent diffusion coefficient and intra-stick diffusivity (Callaghan's model) were fitted and compared between the sequences for glutamate, glutamine (Gln), myo-inositol, taurine, total N-acetylaspartate, total choline, total creatine and the macromolecules.

## Results and discussion

The shorter echo time achieved with DW-SPECIAL (18 ms against 33 ms with STE-LASER) substantially limited the metabolites' signal loss caused by J-evolution. In addition, DW-SPECIAL preserved the main advantages of STE-LASER: absence of cross-terms, diffusion time during a STE and limited sensitivity to $B_1$ inhomogeneities. In vivo, compared to STE-LASER, DW-SPECIAL yielded the same spectral quality and reduced the Cramer Rao Lower Bounds (CRLB) for J-coupled metabolites, irrespective of the b-value. DW-SPECIAL also reduced the standard deviation of the metabolites' diffusion estimates based on individual animal fitting without loss of accuracy compared to the fit on the averaged decay.

## Conclusion

We conclude that due to its reduced echo time, DW-SPECIAL can serve as an alternative to STE-LASER when strongly J-coupled metabolites like Gln are investigated, thereby extending the range of accessible metabolites in the context of DWS acquisitions.

# Introduction

In vivo diffusion-weighted magnetic resonance spectroscopy (DWS) and imaging (DWI) have emerged as powerful techniques to probe tissue morphology at the micrometre scale via the non-invasive assessment of a variety of diffusion metrics[1–4]. By inserting diffusion-sensitizing gradients into traditional single-voxel $^1$H magnetic resonance spectroscopy (MRS) sequences, the diffusion properties of metabolites measured by $^1$H MRS can be retrieved. Investigating the deviation of these diffusion properties from the ones expected for free diffusion allows one to infer the environment that a given metabolite is experiencing.

Contrary to water measured with DWI, brain metabolites are mostly intracellular and some of them are believed to be cell-specific: myo-inositol (mIns) and glutamine (Gln) are predominantly located in the astrocytes and N-acetylaspartate (NAA) and glutamate (Glu) in the neurons[5–7]. Given this prior knowledge, DWS has the potential to provide unique cell-specific microstructural information, synergetic to the non-specific, yet more sensitive information obtained from DWI probing water molecules located in all cell types and exchanging between compartments.

Since the pioneering work in animals and humans[8,9], DWS has explored and revealed microscopic signatures of brain cells[10–13]. Neurons[14], astrocytes[15] and microglia[16] and their alterations in disease populations have been investigated with DWS, also in cases where DWI failed to probe any change due to its non-cell specific nature[15].

However, unlike DWI, DWS suffers from low sensitivity due the low concentration of metabolites compared to water. It is thus important to improve DWS experiments at the acquisition and processing levels.

Traditionally, DWS has been performed using diffusion-weighted (DW-) PRESS or STEAM sequences. Both sequences have their respective advantages and disadvantages: DW-PRESS benefits from full-signal intensity but is impacted by signal losses from the transverse magnetization during the diffusion time due $T_2$ relaxation and J-evolution, and by the limitations of the non-adiabatic 180° pulses at high fields (chemical shift displacement (CSD) artefacts, sensitivity to $B_1$ inhomogeneities with surface coils, high power deposition). DW-STEAM benefits from the slow $T_1$ relaxation of the longitudinal magnetization originating the signal during the diffusion time and from better radio-frequency (RF) pulse selection properties by using only 90° pulses, but has the drawback of the resulting halved signal intensity. Additionally, both DW-STEAM and DW-PRESS are affected by cross-terms, namely contributions to the b-value proportional to $\boldsymbol{g}_{diff} \cdot \boldsymbol{g}_{other}$ (where $\boldsymbol{g}_{other}$ stands for imaging/spoiler/crusher gradients), which need to be accounted for to accurately estimate the diffusion properties.

More recently, the DW-sLASER[17] and the STE-LASER[18] sequences have been introduced, both being based on the LASER[19,20] volume localization. STE-LASER became the gold-standard in rodent DWS studies: its block-design separates the stimulated echo diffusion module from the LASER localization and thus prevents the appearance of cross-terms, while preserving the other advantages of DW-STEAM.

However, the use of STE-LASER leads to longer echo times, thus hindering the detection limits of J-coupled metabolites. Currently, mostly singlets (NAA, total choline (tCho), total creatine (tCr)) and few J-coupled metabolites (taurine (Tau), mIns) are reported in DWS studies. Gln for example is seldom investigated, owing to challenges in its quantification, even more so when the MRS signal is strongly weighted by diffusion: its overlap with Glu at fields < 9.4T, a low concentration and strong J-coupling. Yet, Gln is a desired target for DWS studies as it plays an important role in various pathologies and is an astrocyte-specific marker due to the exclusive location of glutamine synthetase (GS) in the astrocytes[21]. A striking example is hepatic encephalopathy (HE), where brain Gln is increased by more than 100% as a result of excessive ammonia reaching the brain[22–25]. In that context, DWS probed an increase in metabolite diffusivities, including in Gln, in the cerebellum of a rat model of the disease, consistent with the loss of neuronal and astrocytic internal structure observed by histology[26,27]. Yet, a reliable estimation of Gln diffusion properties in the control group, where its concentration is lower, still remains challenging.

The SPECIAL sequence[28] and its semi-adiabatic counterpart[29] have been introduced in animal and human $^1$H MRS studies to reach shorter echo times and thus reduce signal loss by J-evolution and $T_2$ relaxation[30,31]. Following this trend, we propose a new diffusion-weighted MR spectroscopy sequence, the DW-SPECIAL sequence, based on the semi-adiabatic SPECIAL sequence combined with a stimulated echo (STE) diffusion block, with the aim of preserving the advantages of the gold-standard STE-LASER sequence in rodent DWS studies, while reaching a shorter echo time.

## Methods

### Sequence design

The DW-SPECIAL combines a STE diffusion block with a semi-adiabatic SPECIAL[28,29] localization (sequence diagram in **Figure 1**). The first slice-selective 90° pulse is an asymmetric P10 pulse[32] (0.5 ms, 13.5 kHz bandwidth, 3.3 kHz γ$B_{1,max}$, numerically optimized, 5 lobes, 18% refocusing factor), whose gradient refocusing lobe is applied before the first diffusion gradient to avoid cross-terms between these two gradients in the b-value (**Supplementary Materials, Appendix**). Two additional non-slice-selective 90° block pulses (0.1 ms, 12.8 kHz bandwidth) are used to form the STE block. The adiabatic 180° pulses are hyperbolic-secant HS1-R20 pulses[20] (2 ms, 10 kHz bandwidth, 4.8 kHz $\gamma B_{1,max}$ for adiabatic inversion). The slice-selective adiabatic inversion is inserted in the mixing time of the STE block and is applied in the direction with strongest $B_1$ inhomogeneity (here the *y* direction, perpendicular to the transmit/receive (T/R) quadrature surface coil). It is alternatively switched on and off to perform the 1D ISIS scheme (a two-step phase cycling is required to obtain a spectrum). An additional water suppression Hermite pulse (15.4 ms, 350 Hz bandwidth, 89 Hz $\gamma B_{1,max}$) is inserted in the mixing time to saturate the water signal relaxing during the diffusion time. Bipolar diffusion gradients that minimize the effects of eddy currents are used around the two non-slice-selective 90° pulses. Spoiler and crusher gradient amplitudes were adjusted empirically to minimize spurious echoes and outer voxel contributions while limiting the additional unwanted diffusion-weighting created by crushers around the 180° pulses. The following phase cycling was used: $ph_{bp}$ = 0, $ph_{HS,on/off}$ = {0000000022222222}[1], $ph_{P10}$ = {0000222200002222}[1], $ph_{HS}$ = {0011223300112233}[1], $ph_{receive}$ = {0202202002022020}[3] (Bruker's nomenclature: phases in units of 90° in brace brackets, phase increment for the second cycle given by the exponent $n$ ($+n \times 90°$). The corresponding author is willing to share the sequence ready-to-use on Paravision 360 v1.1 and to help adapting it to other versions of Bruker's software and to other field strengths and gradient systems.

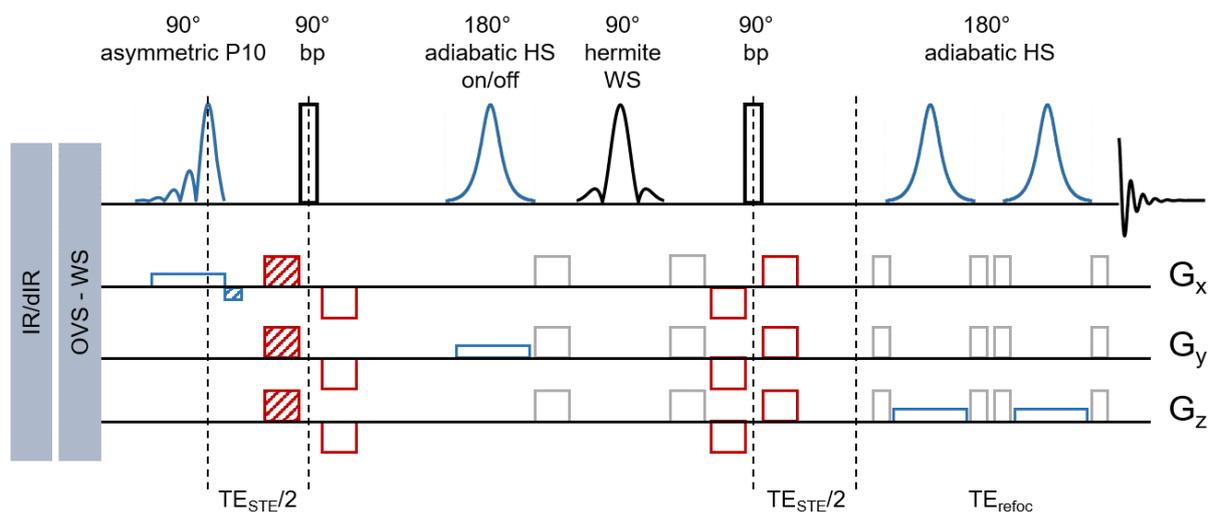

*Figure 1* - DW-SPECIAL sequence. 1st row: RF pulses, the ones from the semi-adiabatic SPECIAL sequence are highlighted in blue. 2nd to 4th row: gradients - Blue: slice-selection gradients, red: bipolar diffusion gradients, grey: spoiler/crusher gradients (arbitrary values displayed). WS, OVS and (d-)IR modules are inserted before the first excitation pulse. The slice-refocusing gradient of the first slice-selective 90° pulse (blue) and the first diffusion gradients (red) are striped to indicate that they should not be applied simultaneously to avoid cross-terms in the b-value calculation. The details of the RF pulses are provided in the **Sequence design** section of the Methods.

## In vivo acquisitions

All experiments were approved by The Committee on Animal Experimentation for the Canton de Vaud, Switzerland (VD3022.1).

Three adult male Wistar rats were scanned twice with a two-week interval to acquire five diffusion sets (rat 1, rat 2 and rat 3 at week 0, rat 1 and rat at week 2) and one macromolecules profile (rat 3 at week 2), with DW-SPECIAL and STE-LASER. During the DWS experiments, isoflurane anaesthesia (~1.5%, in a 50%/50% air/$O_2$ mix, 0.9 L/min) was used, the respiration rate and the body temperature were monitored (SA Instruments, New York, NY, USA), the latter being maintained at 37.7°C by circulating warm water.

Diffusion-weighted $^1$H MRS experiments were performed on a 14.1T Bruker scanner (Avance Neo, Paravision 360 v1.1), with maximum gradient amplitude of 1 T/m (rise time: 270 µs), and a home-made T/R quadrature surface radio-frequency (RF) coil (two loops of 20 mm diameter).

Anatomical MR images were acquired first to position the MRS voxel and define the shimming region using a localizer multi-slice sequence (FOV: 25x25 mm$^2$, 98x98 µm$^2$ in-plane resolution, 15 slices/direction, slice thickness: 0.8 mm, slice gap: 1 mm, TE/TR: 2.92/18 ms, 1 average) and a T2-turbo RARE sequence (FOV: 20x20 mm$^2$, 78x78 µm$^2$ in-plane resolution, 15 axial slices, slice thickness: 1 mm, slice gap: 0.2 mm, rare factor 6, TE/TR: 27/3000 ms, 2 averages).

The position of the MRS voxel (7x5x5 mm$^3$) is displayed in **Figure 2**. First and second order shimming was performed with the Bruker MAPSHIM method (shim values based on a measured map of the $B_0$ field in the object) followed by local iterative first order shimming in the MRS voxel, leading to a water linewidth of 17-19 Hz. For the DW-SPECIAL acquisition, the minimum achievable TE was used (TE = 18.5 ms with $TE_{STE}$ = 8.3 ms, $TE_{refoc}$ = 10.2 ms, as labelled in **Figure 1**), and a mixing time of 40 ms, yielding a diffusion time ($\Delta$) of 42.6 ms (characteristic 1D diffusion length of ≈ 3.5 µm). The STE-LASER sequence[18] was used for comparison with its respective minimum achievable TE of 33.7 ms ($TE_{STE}$ = 8.7 ms, $TE_{refoc}$ = 25 ms) and a mixing time of 40 ms, yielding a diffusion time ($\Delta$) of 43.4 ms.

For both sequences, the diffusion gradients duration (δ) was set to 3 ms. The following b-values in the direction (1,1,1) were used: 0.05, 1, 3, 5, 10 and 30 ms/µm$^2$ for STE-LASER and 0.05, 0.94, 2.8, 4.7, 9.4 and 28.2 ms/µm$^2$ for DW-SPECIAL. The last b-value was acquired in one animal only as a proof of feasibility. The mismatched b-values between DW-SPECIAL and STE-LASER was not intended: for clarity, only the round values from STE-LASER will be referred to in the following text, but the true b-values were used for fitting and display. The following other acquisition parameters were used identically for the two sequences: TR = 3000 ms, 4096 complex points, 7142 Hz of spectral width, carrier frequency for excitation of water-suppressed signals at 3 ppm.

The number of metabolites shots was increased for the last b-value (NS = 160/160/160/160/320) to compensate for the signal to noise ratio (SNR) drop caused by the potentially additional motion-corrupted shots removed during post-processing. The VAPOR water suppression module[32] (pulse bandwidth: 350 Hz, last delay: 22.8 ms, flip angles 1 and 2: 84°/150°) was optimized and interleaved with outer volume suppression (OVS) pulses (6 slabs, slab thickness: 12 mm, gap to voxel: 0.3 mm, sech pulse (1 ms), gradients (x/y/z: 150/250/350 mT/m, 3ms)). A reference non-water suppressed spectrum was acquired for eddy currents correction (ECC) and between-scan $B_0$ drift compensation was performed with a navigator scan. Each b-value was acquired in full as single shots and in a random order between sequences and b-values.

The term *shot* will be used throughout the manuscript to refer to each MRS complex free induction decay (FID) acquired and stored individually (i.e. two shots are needed to complete the ISIS scheme)[33].

## Phantom acquisitions

Phantom experiments were performed to validate the J-evolution observed in the simulated basis-set. Three phantoms were prepared (**a)** 8mM Gln, **b)** 8mM GABA, **c)** 8mM mIns + 8mM Cr, with 0.5mM DSS in PBS) and scanned with the same sequence parameters at b = 0.05 ms/µm$^2$). A diffusion acquisition on a forth phantom containing all the metabolites observed in vivo (see the **Processing** section of the Methods part) with realistic concentrations was also performed. The diffusion attenuation of mIns, Tau, Glu and tCr signals in solution were measured with DW-SPECIAL and STE-LASER to validate experimentally the absence of cross-terms in DW-SPECIAL. The same sequence parameters as for the in vivo acquisitions were used, except for the b-values ranging from 0.05 to 4 ms/µm$^2$.

## Processing

The same processing was applied for the two sequences. Individual shots were corrected for EC with the water signal. Phase and frequency drifts were performed simultaneously with spectral registration in FID-A[34] (time domain, spectral region restricted to the NAA peak at 2.01 ppm, aligned to the median of the shots, 12 Hz line broadening for processing only), followed by motion-corrupted shots removal (mean square error $z_i$ between the spectrum $i$ and the median spectrum with a rejection criterion: $\frac{z_i-\bar{z}}{SD(z)} > 1.5$, where $\bar{z}$ and $SD(z)$ are the mean and SD of $z_i$ across shots). For DW-SPECIAL, the above-mentioned processing was applied separately on odd and even shots and if one shot was corrupted and removed, its corresponding on/off shot from the 1D ISIS scheme was also removed (**Supplementary Materials, Figure S1**). A metabolite basis-set was simulated for each sequence with NMRScope-B[35] (jMRUI[36,37] - http://www.jmrui.eu), using the exact RF pulse shapes and delays. It included: alanine (Ala), ascorbate (Asc), aspartate (Asp), β-hydroxybutyrate (bHB), glycerophosphocholine (GPC), phosphocholine (PCho), creatine (Cr), phosphocreatine (PCr), γ-Aminobutyric acid (GABA), glucose (Glc), Gln, Glu, glutathione (GSH), mIns, lactate (Lac), NAA, Nacetylaspartylglutamate (NAAG), phosphoethanolamine (PE), scyllo-Inositol (Scyllo), and Tau using previously published J-coupling constants and chemical shifts[38–40]. Metabolites concentration were quantified with LCModel and the diffusion coefficients were fitted only for the metabolites with Cramer Rao Lower Bounds (CRLB) below 5% at b = 0.05 ms/µm$^2$ (Glu, mIns, Tau, tNAA, tCr, tCho, and the mobile macromolecules (MM)) and Gln (CRLB = 6.4±0.5%).

The macromolecules displayed in **Figure 2** were acquired in the same voxel in one animal using double inversion-recovery (dIR) and diffusion-weighting[41] (TE = 18.5 ms, TR = 4000 ms, TI = 2200/850 ms for DW-SPECIAL and TI = 2200/800 ms for STE-LASER, 7 blocks of 64 shots, b = 10 ms/µm$^2$), and metabolites residuals were removed using AMARES[42] from jMRUI[43]. The metabolites residual patterns were further confirmed with the acquisition of MM spectra at multiple inversion times and with/without diffusion gradients (**Supplementary Materials, Figure S2**).

A detailed table of the acquisition and processing parameters following the ISMRM experts' consensus recommendations on minimum reporting standards in in vivo MRS (MRSinMRS)[44] is presented in **Supplementary Materials, Table S3.**

### Fitting

A Gaussian diffusion model ($\frac{S}{S_0} = e^{-bADC}$) up to b = 3 ms/µm$^2$ and the randomly-oriented sticks model[45] ($\frac{S}{S_0} = \sqrt{\frac{\pi}{4bD_{intra}}} \, \text{erf}(\sqrt{bD_{intra}})$) up to b = 10 ms/µm$^2$ were fitted to each metabolite concentration decay as a function of b-value. The fits were performed individually for each animal and on the averaged concentration decay normalized to its value at b = 0.05 ms/µm$^2$, the latter being referred to as *mean fit*. A non-linear least squares algorithm was used (Matlab *fit* function, Trust-Region method), weighted with the inverse of absolute CRLB for the individual fit case, and with the standard deviation (SD) of each normalized concentration across animals for the mean fit case.

### Statistics

The apparent diffusion coefficient (ADC) and the intra-stick diffusion coefficient ($D_{intra}$) are reported as mean and SD across animals, and with their corresponding value fitted from the mean decay. We approximated that the two rats scanned twice with a two-week interval and used for the diffusion sets could be considered as independent samples for statistics.

A two-way repeated measures ANOVA was performed on the ADC and $D_{intra}$ values fitted on individual animals' metabolite diffusion decays, comparing the sequence effect for all metabolites, with animal-matched values and Bonferroni multiple comparisons post-hoc test.

For each metabolite ADC or $D_{intra}$, the *mean fit* estimates were also compared to a Gaussian distribution created from the mean and SD of the corresponding individual fits to assess the null hypothesis (mean estimate of individual fit = estimate of the *mean fit*) for a *p*-value of 0.05.

### Results

To compare overall the spectral quality of DW-SPECIAL and STE-LASER, a diffusion set up to b = 10 ms/µm$^2$ was acquired with both sequences. Good and comparable quality between the two sequences was observed at all b-values (**Figure 2C**). **Supplementary Materials**, **Figure S3** also shows that strong diffusion-weighting was feasible (b = 30 ms/µm$^2$) with DW-SPECIAL. Although not shown here, the same quality at b = 30 ms/µm$^2$ was also achieved with STE-LASER[18]. The MM contribution was higher in DW-SPECIAL due to the shorter TE (**Figure 2C**).

To compare the volume selection between the two sequences, the pulse profiles on the three directions were measured in one phantom experiment (**Figure 2B**). Similar *x* and *z* profiles and dimensions were obtained with DW-SPECIAL and STE-LASER (*x*: 6.4 mm for DW-SPECIAL and 6.3 mm for STE-LASER for a nominal size of 7 mm, *z*: 4.8 mm for DW-SPECIAL and 4.6 mm for STE-LASER for a nominal size of 5 mm). The slice-selection profile on *y* shows a higher contribution of signals distant from the coil (towards *y* positive) with STE-LASER than with DW-SPECIAL (4.1 mm and 3.8 mm, respectively), while remaining within the 5 mm nominal selection for both sequences.

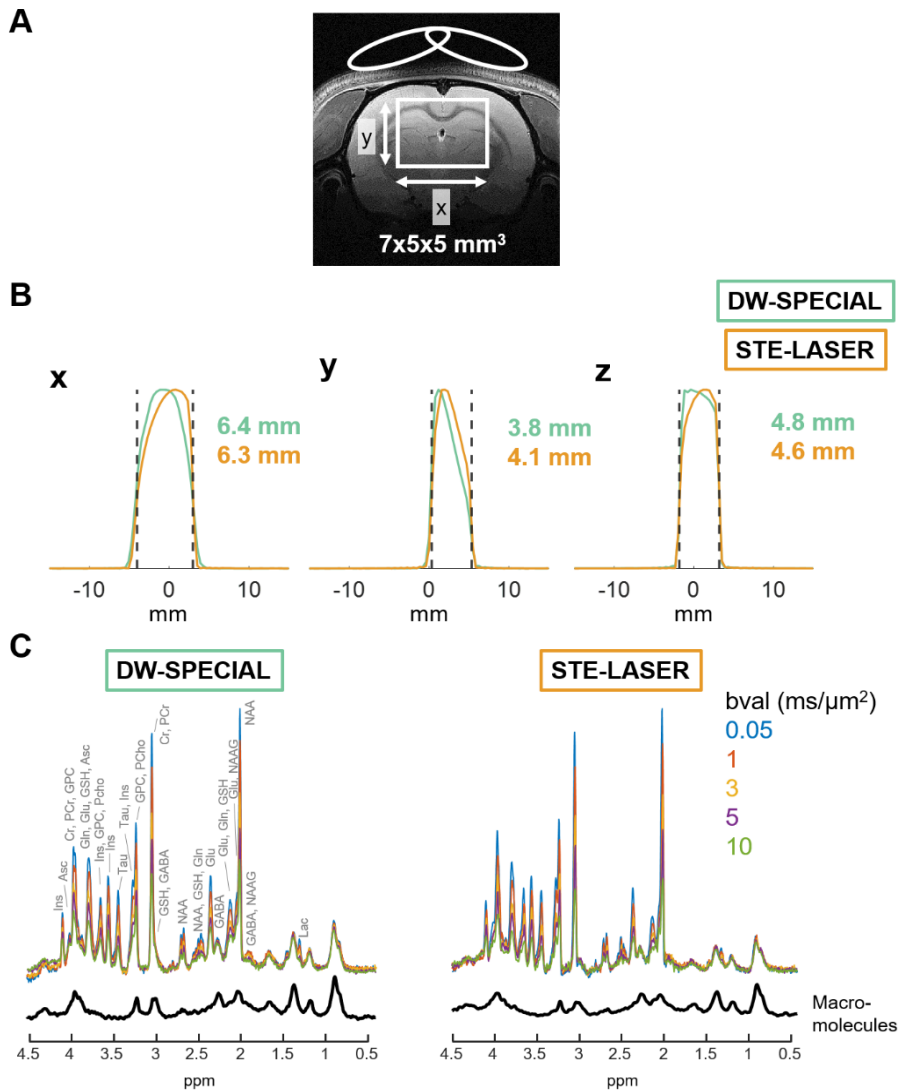

*Figure 2 – A:* Representative voxel location in one animal with x, y directions labelled: y, perpendicular to the plane of the surface coil, is the most inhomogeneous direction in $B_1$ amplitude. *B:* 1D projections of voxel profiles on x, y, z (obtained by switching on a gradient during the acquisition of the water signal) for DW-SPECIAL (green) and STE-LASER (orange), measured in the multi-metabolite phantom with a nominal voxel size of 7x5x5 mm³. The integral values of the profile shapes are displayed. The dashed black lines represent the nominal voxel position in each direction. *C:* Representative in vivo diffusion sets for both sequences, after pre-processing (ECC, phase/frequency drift correction, outlier removal) and 2 Hz line broadening. Macromolecules are also displayed (black). Voxel profiles were very similar on x and z. On y, the $B_1$-inhomogeneous direction, STE-LASER selected more signal distant from the coil (towards y positive). The diffusion sets and macromolecules with both sequences were of good quality.

To validate the J-evolution observed in the basis set simulations (**Figure 3A**), in vitro acquisitions were performed at b = 0.05 ms/µm² with both sequences (**Figure 3B**). The matching J-evolution patterns between the simulations and the in vitro experiments indicated that, for strongly coupled metabolites like Gln, mIns or GABA, the shorter TE achieved in DW-SPECIAL yielded less signal loss due to J-evolution and $T_2$ relaxation.

To confirm experimentally the absence of cross-terms in DW-SPECIAL, the diffusion attenuation of mIns, Tau, Glu and tCr (given as examples) was measured in vitro with both sequences (**Supplementary Materials, Figure S4**) and compared, as STE-LASER is known to have no cross-terms in the b-value. An identical diffusion decay was found for these metabolites, attesting the absence of cross-terms in DW-SPECIAL as well. This was also shown theoretically in **Supplementary Materials, Appendix**.

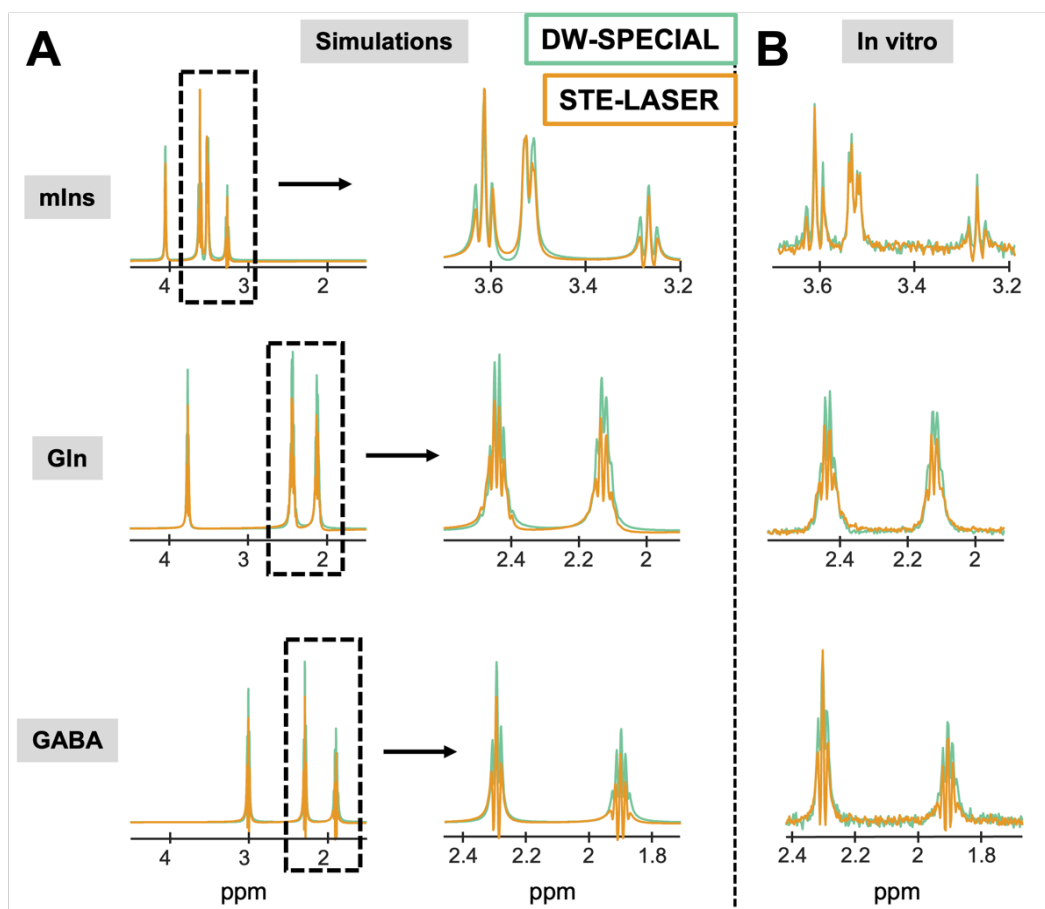

*Figure 3 - Basis set simulations (A) of some selected metabolites (mIns, Gln and GABA) (left column), with a zoom on a multiplet region (central column) and the equivalent spectral region measured in vitro in phantoms (B) for DW-SPECIAL (green) and STE-LASER (orange). Zero-filling and a 5 Hz line broadening were applied for the simulations and a 0, 2 and 5 Hz for mIns, Gln and GABA, respectively, for the in vitro experiments. The simulations were created with the exact delays and pulse shapes for both sequences and accounted for J-evolution but not $T_2$ relaxation. Simulations of Cr for both sequences featured no difference (not displayed here). The multiplet patters measured in vitro matched the simulated ones and confirmed the weaker J-evolution in DW-SPECIAL versus STE-LASER due its shorter total TE.*

To further investigate the spectral quality of single shots for DW-SPECIAL, the pre-processing results with FID-A were analysed and compared to the ones of STE-LASER. The frequency correction factors (**Figure S1C**) were small, stable across b-values and similar between the two sequences (freq$_{corr}$ = 1.2 ± 0.7 Hz for DW-SPECIAL and freq$_{corr}$ = 2.1 ± 1.7 Hz for STE-LASER). The phase correction factors (**Figure S1D**) were increasing with b-value and also comparable between sequences (at b = 0.05 ms/µm$^2$: ph$_{corr}$ = 2.8 ± 0.3° for DW-SPECIAL and $ph_{corr}$ = 3.3 ± 0.6° for STE-LASER, at b = 10 ms/µm$^2$: $ph_{corr}$ = 16.1 ± 3.0° for DW-SPECIAL and $ph_{corr}$ = 24.3 ± 8.5° for STE-LASER). The number of shots removed at each b-value (**Figure S1B**) was higher in DW-SPECIAL versus STE-LASER (at b = 0.05 ms/µm$^2$: $NS_{removed}$ = 29 ± 7 for DW-SPECIAL and $NS_{removed}$ = 11 ± 1 for STE-LASER, at b = 10 ms/µm$^2$: $NS_{removed}$ = 41 ± 14 for DW-SPECIAL and $NS_{removed}$ = 20 ± 9 for STE-LASER) due to the conservative condition of removing the on/off 1D ISIS pair when at least one of the shots is corrupted.

To evaluate the fit quality of the individual animal diffusion decays, LCModel relative CRLB were compared between the sequences. The fit quality improved for DW-SPECIAL compared to STE-LASER for most metabolites, as judged from reduced CRLB, most noticeably for Gln, irrespective of the b-value (**Figure 4B,** Gln: $CRLB_{b3,STE-LASER}$ = 11.2±1.9, $CRLB_{b3,DW-SPECIAL}$ = 7.8±0.8, $CRLB_{b10,STE-LASER}$ = 16.6±4.3, $CRLB_{b10,DW-SPECIAL}$ = 11.6±1.3). Concentration and CRLB tables for all reported metabolites can be found in **Supplementary Materials**, **Table S1** and **S2**.

To evaluate the group homogeneity, the diffusion decays were further averaged across animals for each metabolite and each sequence, after normalization to b = 0.05 ms/µm$^2$. Compared to STE-LASER, DW-SPECIAL improved the within-group homogeneity of diffusion decays (**Figure 4A** and **C**) for J-coupled metabolites like Gln and mIns (Gln: $SD_{b3,STE-LASER}$ = 0.13, $SD_{b3,DW-SPECIAL}$ = 0.02, $SD_{b10,STE-LASER}$ = 0.10, $SD_{b10,DW-SPECIAL}$ = 0.04), while preserving the within-group dispersion obtained with STE-LASER for weakly coupled metabolites (tNAA, MM).

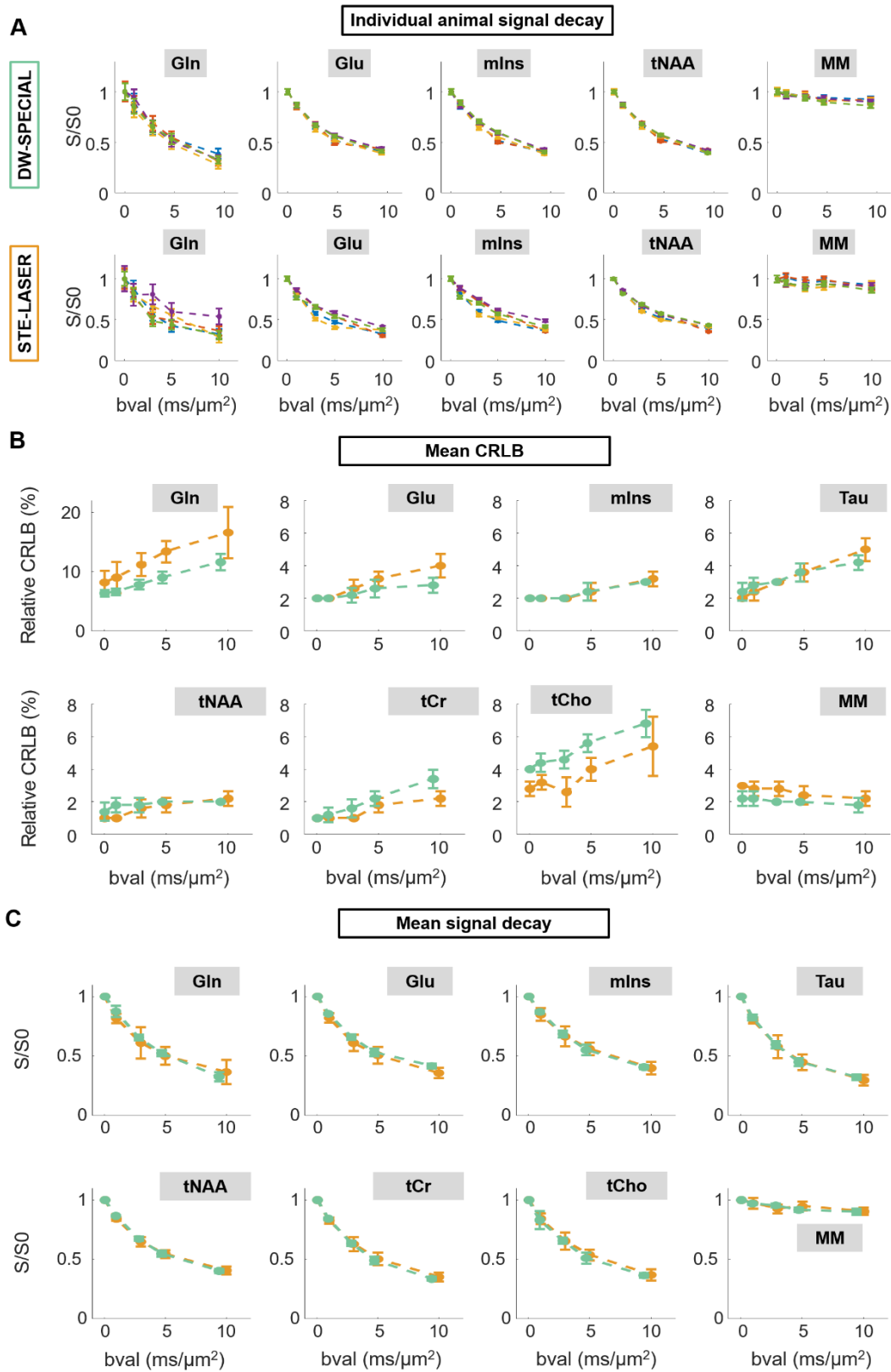

*Figure 4 – A: Concentration decays of Gln, Glu, mIns, tNAA and MM as a function of b-value for all animals (different colors) and both sequences, normalized to their value at b = 0.05 ms/µm². Error bars: absolute CRLB from LCModel quantification. B: Relative CRLB (%) averaged over animals, as a function of b-value, for both sequences (DW-SPECIAL: green, STE-LASER: orange) and all quantified metabolites. Error bars: SD across animals at each b-value. For the relative CRLB, all metabolites are plotted with the same y-range except Gln. Of note, LCModel relative CRLB output being rounded to the nearest integer %, the SD for the CRLB are artificially high. C: Normalized concentration decays averaged over animals, as a function of b-value, for both*

*sequences (DW-SPECIAL: green, STE-LASER: orange) and all quantified metabolites. DW-SPECIAL improved the within-group homogeneity of diffusion decays and improved or equalled LCModel fit quality (reduced relative CRLB) with respect to STE-LASER, for all metabolites except for tCho and tCr. For these two metabolites, the relative CRLB still remained low at all b-values for the two sequences (below 8% for tCho and below 4% for tCr).*

Finally, to assess the reliability of the diffusion estimates derived from DW-SPECIAL, the ADC and $D_{intra}$ from individual animal decays and from the mean decay were compared between the sequences. DW-SPECIAL reduced the SD of estimated ADC and $D_{intra}$ between animals (expected to be part of a homogenous control cohort) (**Figure 5A** and **B,** Gln: $SD_{ADC,STE-LASER}$ = 0.073 µm²/ms, $SD_{ADC,DW-SPECIAL}$ = 0.013 µm²/ms, $SD_{Dintra,STE-LASER}$ = 0.25 µm²/ms, $SD_{Dintra,DW-SPECIAL}$ = 0.076 µm²/ms). No significant difference (ADC: *p*-value = 0.4, $D_{intra}$: *p*-value = 0.9) was found between the two sequences for the mean ADC or $D_{intra}$ of the individual metabolite fits. For DW-SPECIAL, the mean fit for all metabolites was also not significantly different from the mean of ADC or $D_{intra}$ of the individual animal fits when assuming a Gaussian distribution around the mean and SD across animals (p > 0.05, **Figure 5C** and **D**).

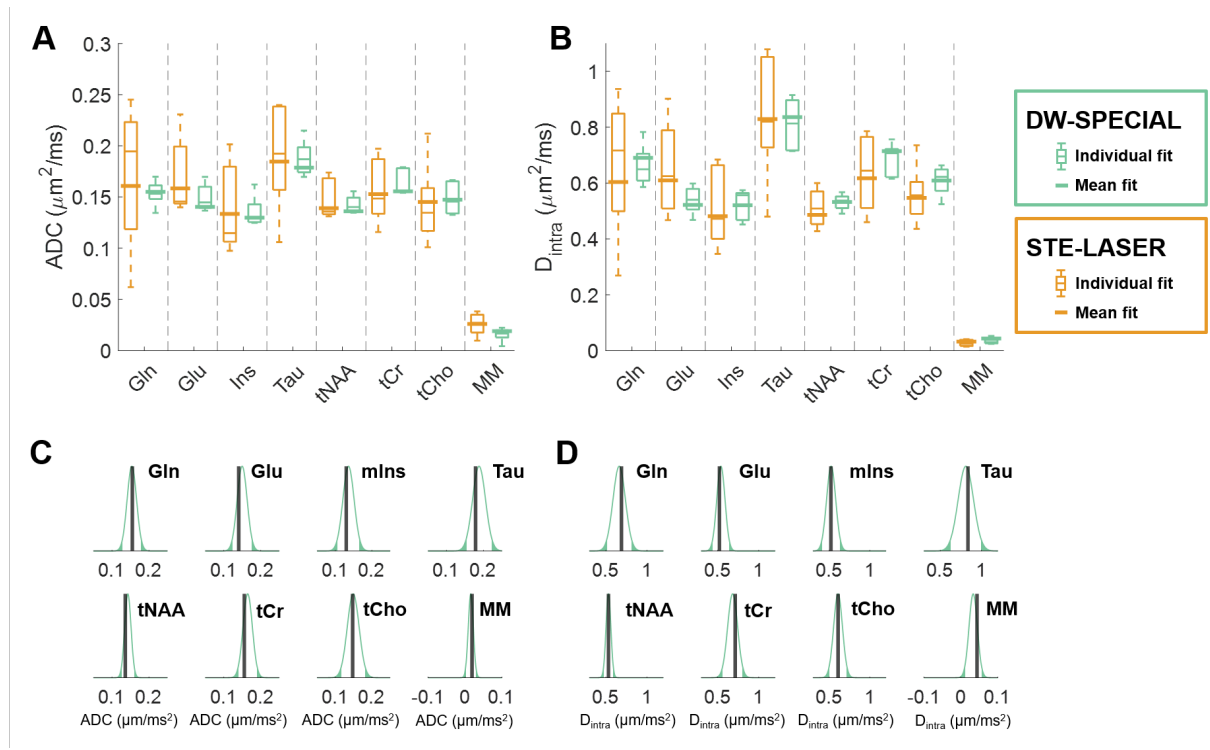

*Figure 5 – ADC (**A**) and $D_{intra}$ (**B**) fitted for all animals with both sequences. Box plots: individual fit for each animal (line: median, top and bottom edges: 25th and 75th percentiles, whiskers: extreme values, dots: outliers). Wide bar: fitted ADC and $D_{intra}$ on the averaged concentration decay over all animals ("mean fit"). $D_{intra}$ from the mean fit was in very good agreement between the two sequences and the SD across animals were reduced with DW-SPECIAL for all metabolites, most notably for the J-coupled ones for which the improvement is major. No statistically significant difference was found for the individually fitted ADC or $D_{intra}$ for any of the metabolites between the two sequences (p-value = 0.4 for ADC, p-value = 0.9 for $D_{intra}$, two-way repeated measures ANOVA). **C** and **D**: from the mean and SD of individual fits with DW-SPECIAL, Gaussian distributions of ADC (panel **C**) and $D_{intra}$ (panel **D**) were generated for each metabolite and compared to the mean fit (wide black bar) (p<0.025 regions on each side of the Gaussian distribution are dark green). The mean fit fell into the Gaussian distribution of the individually fitted ADC and $D_{intra}$ (p>0.05) for all metabolites, which confirmed the agreement between the estimates from the mean fit and from the individual fits for DW-SPECIAL.*

## Discussion

This paper reports the first implementation and validation of a new diffusion-weighted MRS sequence, the DW-SPECIAL sequence. Our goal was to preserve the advantages of STE-LASER[18] used for rodent DWS acquisitions (absence of cross-terms, slower $T_1$ than $T_2$ relaxation during the diffusion time, good voxel localization with limited CSD artefacts and limited sensitivity to $B_1$ inhomogeneities) while reaching a shorter echo time. By doing so, we improved the detection and estimation of diffusion metrics of J-coupled metabolites and widened the range of investigated metabolites in rodent high field DWS studies (e.g. to Gln, a metabolite of interest in numerous pathologies).

### Preserved advantages of the STE-LASER sequence

Our initial goal was to create a new sequence that will benefit from the same advantages of STE-LASER over other conventional DWS sequences such as diffusion-weighted STEAM, diffusion-weighted PRESS or diffusion-weighted semi-LASER, with a shorter TE. Firstly, DW-SPECIAL allows one to reach long diffusion times, the diffusion gradients being inserted in a stimulated echo block where the magnetization relaxes with $T_1$ instead of $T_2$ (like STE-LASER and DW-STEAM but unlike DW-PRESS and DW-LASER). Secondly, asymmetric 90° and adiabatic 180° pulses led to a sharp volume selection and a limited sensitivity to $B_1$ inhomogeneities created by a surface coil (like DW-LASER and DW-STEAM but unlike DW-PRESS). The slice-selection profile of the asymmetric P10 pulse[32] along the *x* dimension matched well the one generated by the two adiabatic 180° pulses used in STE-LASER (**Figure 2B**). Its large bandwidth at lower power compared to a symmetric 90° pulse limited CSD artefacts at ultra-high field[46] and its slice-selective nature limited the presence of spurious echoes originating from outside the VOI, where STE-LASER excites the whole volume before performing 3D localization[31]. The *y* profiles for both sequences are convoluted with the coil sensitivity drop at reception on the axis perpendicular to the surface coil, but the smaller contribution of signals distant from the coil with DW-SPECIAL did not substantially reduce the selected *y* dimension. The *y* profile in DW-SPECIAL was also similar to the profile shown in the SPECIAL sequence original paper[28]. The *z* profile is expected to be similar for DW-SPECIAL and STE-LASER as this dimension is selected by a pair of adiabatic pulses in each case: the remaining difference could originate from the shift of the *x* and *y* profiles' maximum between the two sequences, leading to a contribution of signals experiencing different effective $B_1$ fields. To validate the voxel selection with DW-SPECIAL, the 3D profiles were measured for a smaller voxel in the same phantom (3x3x3 mm$^3$) positioned in a $B_1$-homogeneous region. In that case, almost identical profiles were observed between DW-SPECIAL and STE-LASER for the three directions (**Supplementary Materials, Figure S5**), further confirming the accurate volume selection with DW-SPECIAL.

Finally, an attractive aspect of the STE-LASER sequence is its block-design, where the diffusion weighting is fully separated from the localization performed with the LASER sequence, leading to the absence of cross-terms between diffusion and imaging gradients contributing to the b-value (unlike DW-STEAM, DW-PRESS and DW-LASER). Although this block-design was not preserved in our newly proposed sequence, the absence of cross-terms was however ensured as follows: **a)** the localization gradient applied during the on/off 180° ISIS pulse in the mixing time does not take part in the b-value calculation (like all other gradients in the mixing time[47]), and **b)** the slice-refocusing gradient of the first slice-selective 90° P10 pulse and the first diffusion gradient (striped in **Figure 1**) were not applied simultaneously to prevent cross-terms originating from the first part of the STE. The absence of cross-terms was supported experimentally with the within error in vitro diffusion attenuations of mIns, Tau, Glu and tCr with DW-SPECIAL and STE-LASER and a detailed calculation of the b-value (**Supplementary Materials, Appendix** and **Figure S4**). To minimize the increase in minimum TE caused by the separation of these two gradients, we used an asymmetric 90° pulse with an 18% refocusing factor, thus limiting the slice-selective gradient duration and contribution to the echo time. The echo time of the STE diffusion period was similar between the two sequences and mostly governed by the duration of the diffusion gradients. The shorter total echo time achieved with DW-SPECIAL arose from the use of one pair of adiabatic pulses after the STE block, instead of three with STE-LASER.

The pre-processing yielded similar efficiency between the two sequences (**Supplementary Materials**, **Figure S1**), as measured by the amplitude of the frequency and phase correction factors, confirming similar data quality. The frequency-drift correction was small and consistent across b-values ascribed to the recording of a navigator scan at the end of each TR used to dynamically adjust the $B_0$ frequency between each acquisition. The phase-drift, however, increased with b-value due to gradient imperfections, motion and flow. Although the number of shots removed did not impair the spectral quality of DW-SPECIAL, a less conservative condition for outlier removal could be considered: indeed, instead of removing an ISIS on/off pair as soon as at least one of the shots is corrupted, one could equalize, after outlier identification on each of the two subsets, the number of odd and even shots over the total number of shots, irrespective of the pairs forming each ISIS module.

### Improved detection of J-coupled metabolites

In addition to the forementioned properties, the DW-SPECIAL sequence allows one to nearly halve the minimum echo time as compared to STE-LASER (18 ms versus 33 ms). As predicted by simulations, this led to an improved detection of J-coupled metabolites, such as Gln, mIns or GABA experimentally (in phantoms (**Figure 3**) and in vivo (**Figure 4** and **Figure 5**)). To ensure a reliable comparison of the diffusion properties obtained at different echo times (here with two different sequences), there should be no correlation between the metabolites' relaxation and diffusion properties. This aspect has been investigated in vivo in the mouse brain at 11.7T[48] and showed negligible correlation between the metabolites' relaxation and diffusion properties for a range of echo times between 50 to 110 ms. We expect that this observation can be extended to the range of echo times used in the present work (18 to 33 ms), at least for intracellular metabolites, thus ensuring a reliable comparison of the two sequences.

Although the LASER sequence benefits from a reduced apparent J-evolution and $T_2$ relaxation compared to other single-voxel spectroscopy sequences at the same TE due to the succession of adiabatic 180° pulses[49,50] and its current implementation could be further optimized[51], the SPECIAL localization is advantageous when short echo times are desired[30,31]. At lower fields, the echo time could be even further reduced by converting the pair of adiabatic pulses into a single large bandwidth asymmetric 180° (i.e. converting the semi-adiabatic SPECIAL an asymmetric SPECIAL, as initially proposed[28]). The improvement brought by shortening the echo time was particularly clear for Gln, as shown by a better group homogeneity of diffusion decays (left-most panels of **Figure 4A**) and a better LCModel fit quality (CRLB, top left panel of **Figure 4B**) with DW-SPECIAL. An improved Gln quantification with shorter TE could be directly beneficial as Gln concentration is low in healthy cohorts. In hepatic encephalopathy for example, brain Gln can be elevated by more than 100% in rodents[22] and is thus well quantified in the disease group, but remains low in the control group, where DW-SPECIAL could help better quantifying its diffusion properties. Importantly, although well quantified with both sequences, the within-group dispersion of Glu diffusion decays was reduced with DW-SPECIAL, possibly due to a better quantification of Gln and thus a better separation of Glx (Gln + Glu) into Gln and Glu. GABA is also strongly J-coupled and is generally not reported in diffusion studies due to its low concentration and difficult spectral resolution. Data quality obtained with DW-SPECIAL may facilitate the quantification of the diffusion decays of such low-concentrated metabolites (**Supplementary Materials, Figure S6**) and/or the access to higher b-values. For the metabolites traditionally reported in rodent DWS studies (NAA, tCr, tCho, Glu, mIns, Tau), the ADC values were in good agreement with the ones measured in the mouse brain at 11.7T[15], slightly higher in the present study due to the shorter diffusion time.

Another important feature of DW-SPECIAL is that it may render possible the fit of ADC and $D_{intra}$ based on individual animal diffusion decays. Due to the low SNR of DWS experiments, the authors in the DWS community have traditionally fitted the diffusion coefficients on the normalized signal decay averaged over animals or subjects, or on cohort-averaged spectra. These two approaches render the estimation of the uncertainty difficult. Even when error propagation or Monte Carlo simulations are used, the computed error on the diffusion coefficients may not accurately represent the dispersion within one group. The agreement between the mean fit and the fit of individual animals for DW-SPECIAL (**Figure 5C** and **D**) suggests that, with the spectral quality obtained herein, individual animal fitting may become feasible without a substantial penalty in accuracy.

Although the choice of the diffusion model is outside the scope of the present manuscript, it should be noted that the randomly oriented sticks model may not apply well to the macromolecules diffusion decay, which can be described by a mono-exponential decay up to high b-values[52].

### Translation to human scanners and limitations

The implementation of DW-SPECIAL in human scanners is feasible due to its lower number of RF pulses compared to STE-LASER, as LASER-based sequences result in an elevated specific absorption rate (SAR), especially at ultra-high fields[31]. In DW-SPECIAL, two pairs of 180° adiabatic pulses were replaced by an on/off adiabatic 180° pulse and an asymmetric 90° pulse, which considerably reduced the SAR. However, the on/off 1D ISIS module in DW-SPECIAL makes it sensitive to motion artefacts. Whereas this is not a problem in general for rodent experiments where the animal head is fixed by stereotaxic bars, additional care should be taken in human experiments where localized RF calibration and the use of OVS are recommended[31].

## Conclusion

We conclude that the reduced echo time achieved in DW-SPECIAL improves the detection of short $T_2$ and J-coupled metabolites compared to STE-LASER, the current gold-standard in rodent DWS studies at high fields, while preserving the absence of cross-terms in the b-value. Taken together, these advantages make DW-SPECIAL a good candidate to extend the range of investigated metabolites, e.g. Gln, which is rarely reported in DWS studies. We further conclude that the reduced number of RF pulses makes DW-SPECIAL an attractive alternative for human DWS studies, especially at high fields.

## Acknowledgments

Supported by the European Union's Horizon 2020 research and innovation program under the Marie Sklodowska-Curie grant agreement No 813120 (INSPiRE-MED), the SNSF projects no 310030_173222, 310030_201218 and the Leenaards and Jeantet Foundations. We acknowledge access to the facilities and expertise of the CIBM Center for Biomedical Imaging, founded and supported by Lausanne University Hospital (CHUV), University of Lausanne (UNIL), Ecole polytechnique fédérale de Lausanne (EPFL), University of Geneva (UNIGE) and Geneva University Hospitals (HUG). We acknowledge Stefanita Mitrea and Dario Sessa for the BDL surgeries, Analina Da Silva and Mario Lepore for veterinary support, Thanh Phong Lê for technical support, Vladimir Mlynarik for experimental advice and Katarzyna Pierzchala for her help with phantom preparation.

## List of figures:

**Figure 1 -** W-SPECIAL sequence. 1st row: RF pulses, the ones from the semi-adiabatic SPECIAL sequence are highlighted in blue. 2nd to 4th row: gradients - Blue: slice-selection gradients, red: bipolar diffusion gradients, grey: spoiler/crusher gradients (arbitrary values displayed). WS, OVS and (d-)IR modules are inserted before the first excitation pulse. The slice-refocusing gradient of the first slice-selective 90° pulse (blue) and the first diffusion gradients (red) are striped to indicate that they should not be applied simultaneously to avoid cross-terms in the b-value calculation. The details of the RF pulses are provided in the **Sequence design** section of the Methods.

**Figure 2** – **A:** Representative voxel location in one animal with x, y directions labelled: y, perpendicular to the plane of the surface coil, is the most inhomogeneous direction in $B_1$ amplitude. **B:** 1D projections of voxel profiles on x, y, z (obtained by switching on a gradient during the acquisition of the water signal) for DW-SPECIAL (green) and STE-LASER (orange), measured in the multi-metabolite phantom with a nominal voxel size of 7x5x5 mm$^3$. The integral values of the profile shapes are displayed. The dashed black lines represent the nominal voxel position in each direction. **C:** Representative in vivo diffusion sets for both sequences, after pre-processing (ECC, phase/frequency drift correction, outlier removal) and 2 Hz line broadening. Macromolecules are also displayed (black). Voxel profiles were very similar on x and z. On y, the $B_1$-inhomogeneous direction, STE-LASER selected more signal distant from the coil (towards y positive). The diffusion sets and macromolecules with both sequences were of good quality.

**Figure 3 -** Basis set simulations **(A)** of some selected metabolites (mIns, Gln and GABA) (left column), with a zoom on a multiplet region (central column) and the equivalent spectral region measured in vitro in phantoms **(B)** for DW-SPECIAL (green) and STE-LASER (orange). Zero-filling and a 5 Hz line broadening were applied for the simulations and a 0, 2 and 5 Hz for mIns, Gln and GABA, respectively, for the in vitro experiments. The simulations were created with the exact delays and pulse shapes for both sequences and accounted for J-evolution but not $T_2$ relaxation. Simulations of Cr for both sequences featured no difference (not displayed here). The multiplet patters measured in vitro matched the simulated ones and confirmed the weaker J-evolution in DW-SPECIAL versus STE-LASER due its shorter total TE.

**Figure 4** – **A:** Concentration decays of Gln, Glu, mIns, tNAA and MM as a function of b-value for all animals (different colors) and both sequences, normalized to their value at b = 0.05 ms/µm$^2$. Error bars: absolute CRLB from LCModel quantification. **B:** Relative CRLB (%) averaged over animals, as a function of b-value, for both sequences (DW-SPECIAL: green, STE-LASER: orange) and all quantified metabolites. Error bars: SD across animals at each b-value. For the relative CRLB, all metabolites are plotted with the same y-range except Gln. Of note, LCModel relative CRLB output being rounded to the nearest integer %, the SD for the CRLB are artificially high. **C:** Normalized concentration decays averaged over animals, as a function of b-value, for both sequences (DW-SPECIAL: green, STE-LASER: orange) and all quantified metabolites. DW-SPECIAL improved the within-group homogeneity of diffusion decays and improved or equalled LCModel fit quality (reduced relative CRLB) with respect to STE-LASER, for all metabolites except for tCho and tCr. For these two metabolites, the relative CRLB still remained low at all b-values for the two sequences (below 8% for tCho and below 4% for tCr).

**Figure 5** – ADC (**A**) and $D_{intra}$ (**B**) fitted for all animals with both sequences. Box plots: individual fit for each animal (line: median, top and bottom edges: 25$^{th}$ and 75$^{th}$ percentiles, whiskers: extreme values, dots: outliers). Wide bar: fitted ADC and $D_{intra}$ on the averaged concentration decay over all animals ("mean fit"). $D_{intra}$ from the mean fit was in very good agreement between the two sequences and the SD across animals were reduced with DW-SPECIAL for all metabolites, most notably for the J-coupled ones for which the improvement is major. No statistically significant difference was found for the individually fitted ADC or $D_{intra}$ for any of the metabolites between the two sequences (p-value = 0.4 for ADC, p-value = 0.9 for $D_{intra}$, two-way repeated measures ANOVA). **C** and **D**: from the mean and SD of individual fits with DW-SPECIAL, Gaussian distributions of ADC (panel **C**) and $D_{intra}$ (panel **D**) were generated for each metabolite and compared to the mean fit (wide black bar) (p<0.025 regions on each side of the Gaussian distribution are dark green). The mean fit fell into the Gaussian distribution of the individually fitted ADC and $D_{intra}$ (p>0.05) for all metabolites, which confirmed the agreement between the estimates from the mean fit and from the individual fits for DW-SPECIAL.

## List of figures and table – Supplementary information:

**Figure S1** – Processing results for the two sequences. **A**: visualisation of the processing results for one spectrum at b = 0.05 ms/µm$^2$ acquired with DW-SPECIAL. **B**: Number of motion-corrupted shots removed for each sequence, 1 dot per animal and b-value, one color per b-

value. **C**: Frequency correction factors (absolute value) found by spectral registration in FID-A, mean and SD across shots, 1 error-bar per animal and b-value, one color per b-value. **D**: Phase correction factors (absolute value) found by spectral registration in FID-A, mean and SD across shots, 1 error-bar per animal and b-value, one color per b-value. The frequency and phase factors are displayed in absolute value but all the distributions are centred on 0. Due to the conservative condition of removing the on/off 1D ISIS pair when at least one of the shots is corrupted, the number of shots removed at each b-value is higher in DW-SPECIAL versus STE-LASER. Yet, similar phase and frequency correction factors are obtained with both sequences, confirming the good data quality of DW-SPECIAL.

**Figure S2** – Validation of metabolite residuals removal on the macromolecule spectrum using multiple double inversion recovery experiments (**A**) and with high/low diffusion-weighting (**B**). In **A**, the first TI delay was fixed to 2200 ms and the second TI delay was varied (colors) to identify the metabolites residuals, also following Cudalbu et al. and Simicic et al. (https://doi.org/10.1002/nbm.4393 and https://doi.org/10.1002/mrm.28910). In **B**, the second TI delay was fixed to 850 ms (the one featuring the least metabolite residuals, the one chosen in the present manuscript) and the macromolecular spectrum was acquired with high/low diffusion-weighting to confirm the pattern of metabolites removal found in **A**. For visual inspection, the amplitude of the MM at 0.9ppm was matched for the two conditions.

**Figure S3** – Representative DW-SPECIAL diffusion spectra after processing (ECC, phase and frequency alignment and motion-corrupted shots removal), for four b-values, plotted with 5 Hz line broadening. Metabolites with relative CRLB below 10% at $b = 0.05$ ms/µm$^2$ are labelled on the first panel. The spectrum at $b = 30$ ms/µm$^2$ acquired in one animal is shown here as a proof of feasibility. The diffusion spectra are of good quality and the increasing contribution of the macromolecules with respect to the metabolites can be observed with increasing b-values as a result of their slower diffusivity.

**Figure S4** – Phantom experiment confirming the absence of cross-terms in DW-SPECIAL. A phantom mimicking realistic in vivo metabolite concentrations (left panel, 4 Hz line broadening for visualisation) was scanned with DW-SPECIAL and STE-LASER, the latter featuring no cross-terms in the b-value as previously shown in Ligneul et al. (https://doi.org/10.1002/mrm.26217). The example diffusion attenuations over b-value (after normalization to the first b-value) for mIns, Tau, Glu, and tCr after LCModel quantification show no difference between the two sequences, confirming the absence of cross terms in DW-SPECIAL as well. The small offset observed for mIns could have been caused by a quantification error on the first point for either one of the sequences.

**Figure S5** – 1D projections of the x, y, z selection profiles for a small voxel (3x3x3 mm$^3$) in a B$_1$-homogeneous region measured with DW-SPECIAL (green) and STE-LASER (orange) in the multi-metabolite phantom. The profiles are obtained by switching on a gradient during the acquisition. The integral values of the profile shapes are displayed. The dashed black lines represent the nominal voxel position in each direction. Negligible difference is observed in the voxel selection between the two sequences when the factor of B$_1$ inhomogeneity is removed, confirming the good selection performed with DW-SPECIAL.

**Figure S6** – Concentration decays of Asc, Lac, GSH and GABA as a function of b-value for all animals (different colors) and both sequences, normalized to their value at b = 0.05 ms/µm$^2$. Error bars: absolute CRLB from LCModel quantification. Although these metabolites are generally not reported in DWS studies owing to their low concentration and poor quantification, the shorter TE achieved in DW-SPECIAL leads to a smaller within-group dispersion of their diffusion decays and suggests that these metabolites could possibly be investigated in future studies (through a fit of the mean diffusion decay).

**Table S1** - Concentrations and relative Cramer Rao Lower Bounds from LCModel fit of DW-SPECIAL spectra averaged over animals at all b-values and for every reported metabolite. Metabolite concentrations are not referenced to water and their concentrations are in arbitrary unit: only their relative contributions at b = 0.05 ms/µm$^2$ is of biological relevance. The MM basis-set spectrum was scaled such that its concentration at b = 0.05 ms/µm$^2$ is in the range 1-4mM (Cudalbu et al., https://doi.org/10.1002/nbm.4393).

**Table S2** - Concentrations and relative Cramer Rao Lower Bounds from LCModel fit of STE-LASER spectra averaged over animals at all b-values and for every reported metabolite. Metabolite concentrations are not referenced to water and their concentrations are in arbitrary unit: only their relative contributions at b = 0.05 ms/µm$^2$ is of biological relevance. The MM basis-set spectrum was scaled such that its concentration at b = 0.05 ms/µm$^2$ is in the range 1-4mM (Cudalbu et al., https://doi.org/10.1002/nbm.4393).

**Table S3** - MRSinMRS checklist from Lin et al. « Minimum Reporting Standards for in Vivo Magnetic Resonance Spectroscopy (MRSinMRS): Experts' Consensus Recommendations ». NMR in Biomedicine 34, no 5 (2021)

**Diffusion-weighted SPECIAL improves the detection of J-coupled metabolites at high magnetic field**

Corresponding author: Jessie Mosso (jessie.mosso@epfl.ch)

# Supplementary materials

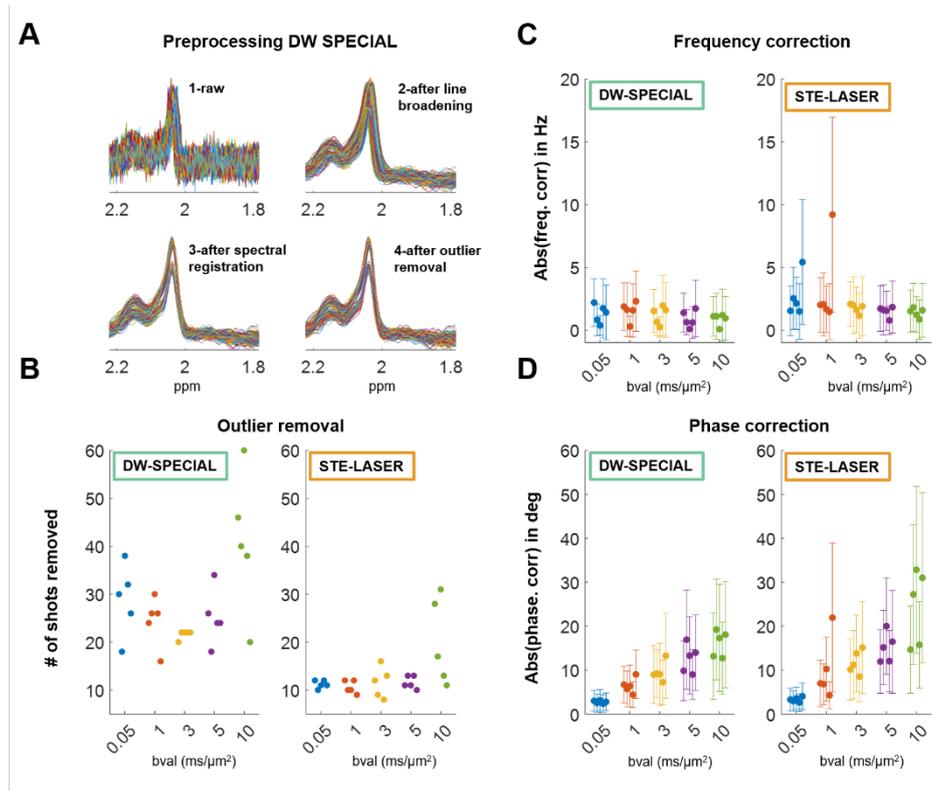

*Figure S1* – *Processing results for the two sequences. **A**: visualisation of the processing results for one spectrum at b = 0.05 ms/µm$^2$ acquired with DW-SPECIAL. **B**: Number of motion-corrupted shots removed for each sequence, 1 dot per animal and b-value, one color per b-value. **C**: Frequency correction factors (absolute value) found by spectral registration in FID-A, mean and SD across shots, 1 error-bar per animal and b-value, one color per b-value. **D**: Phase correction factors (absolute value) found by spectral registration in FID-A, mean and SD across shots, 1 error-bar per animal and b-value, one color per b-value. The frequency and phase factors are displayed in absolute value but all the distributions are centred on 0. Due to the conservative condition of removing the on/off 1D ISIS pair when at least one of the shots is corrupted, the number of shots removed at each b-value is higher in DW-SPECIAL versus STE-LASER. Yet, similar phase and frequency correction factors are obtained with both sequences, confirming the good data quality of DW-SPECIAL.*

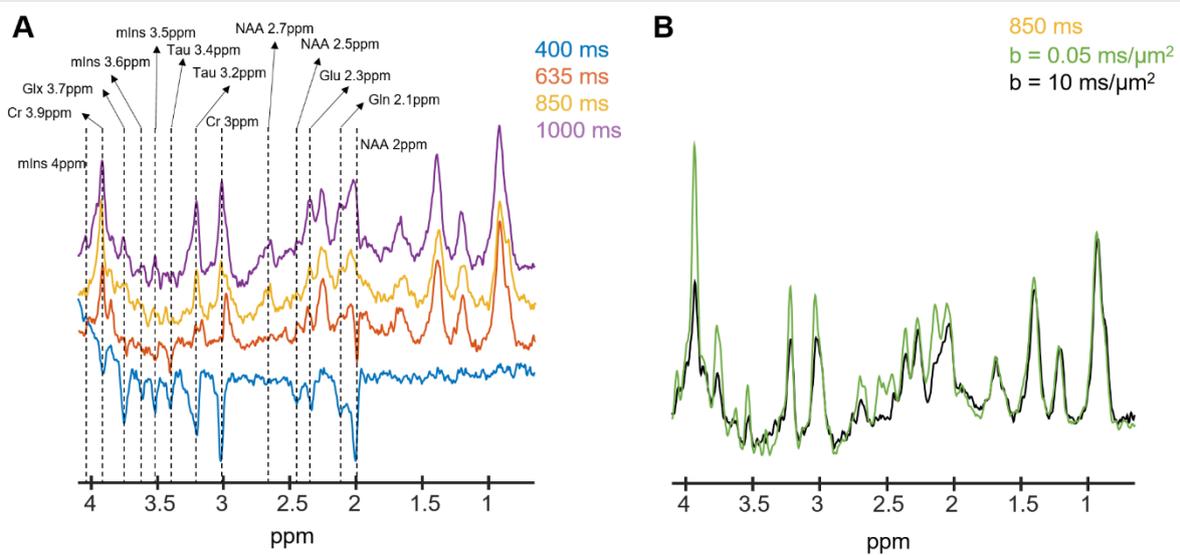

*Figure S2* – Validation of metabolite residuals removal on the macromolecule spectrum using multiple double inversion recovery experiments (**A**) and with high/low diffusion-weighting (**B**). In **A**, the first TI delay was fixed to 2200 ms and the second TI delay was varied (colors) to identify the metabolites residuals, also following Cudalbu et al. and Simicic et al. (https://doi.org/10.1002/nbm.4393 and https://doi.org/10.1002/mrm.28910). In **B**, the second TI delay was fixed to 850 ms (the one featuring the least metabolite residuals, the one chosen in the present manuscript) and the macromolecular spectrum was acquired with high/low diffusion-weighting to confirm the pattern of metabolites removal found in **A**. For visual inspection, the amplitude of the MM at 0.9ppm was matched for the two conditions.

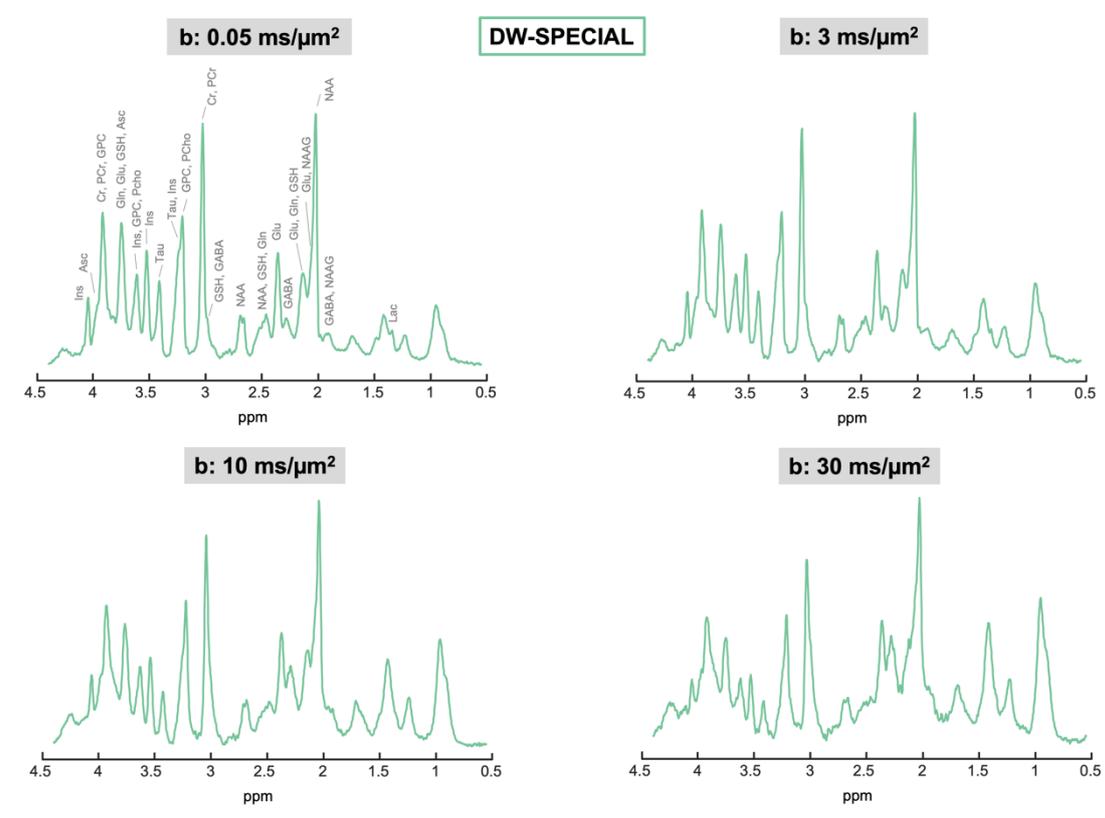

*Figure S3* – Representative DW-SPECIAL diffusion spectra after processing (ECC, phase and frequency alignment and motion-corrupted shots removal), for four b-values, plotted with 5 Hz line broadening. Metabolites with relative CRLB below 10% at b = 0.05 ms/µm² are labelled on the first panel. The spectrum at b = 30 ms/µm² acquired in one animal is shown here as a proof

*of feasibility. The diffusion spectra are of good quality and the increasing contribution of the macromolecules with respect to the metabolites can be observed with increasing b-values as a result of their slower diffusivity.*

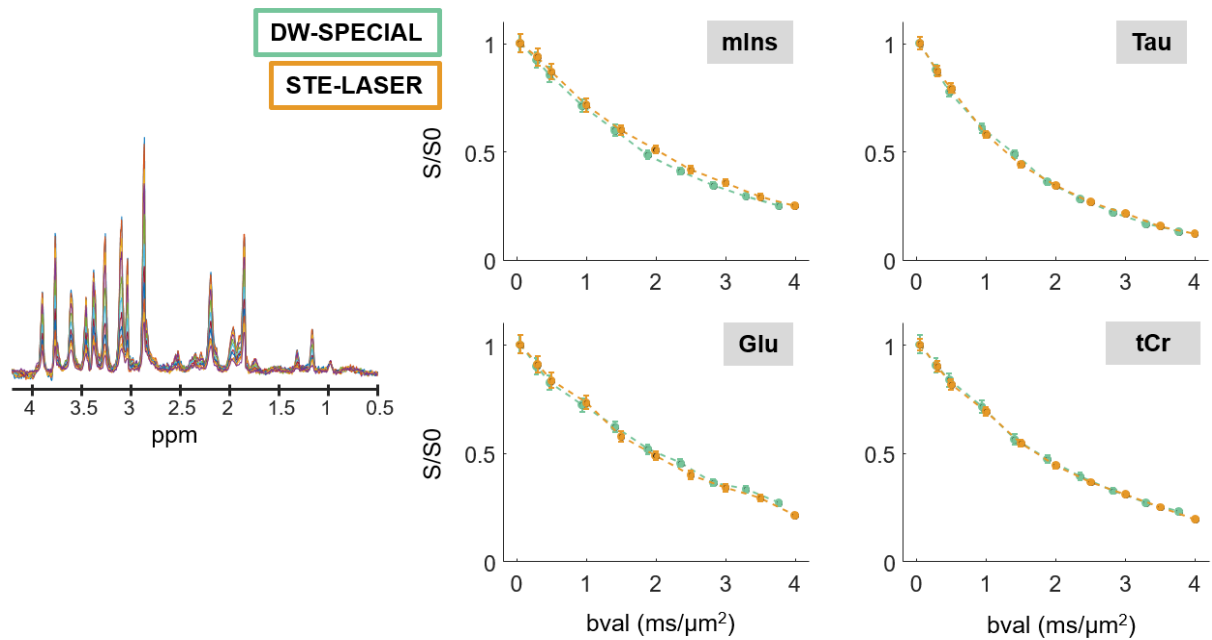

*Figure S4 – Phantom experiment confirming the absence of cross-terms in DW-SPECIAL. A phantom mimicking realistic in vivo metabolite concentrations (left panel, 4 Hz line broadening for visualisation) was scanned with DW-SPECIAL and STE-LASER, the latter featuring no cross-terms in the b-value as previously shown in Ligneul et al. (https://doi.org/10.1002/mrm.26217). The example diffusion attenuations over b-value (after normalization to the first b-value) for mIns, Tau, Glu, and tCr after LCModel quantification show no difference between the two sequences, confirming the absence of cross terms in DW-SPECIAL as well. The small offset observed for mIns could have been caused by a quantification error on the first point for either one of the sequences.*

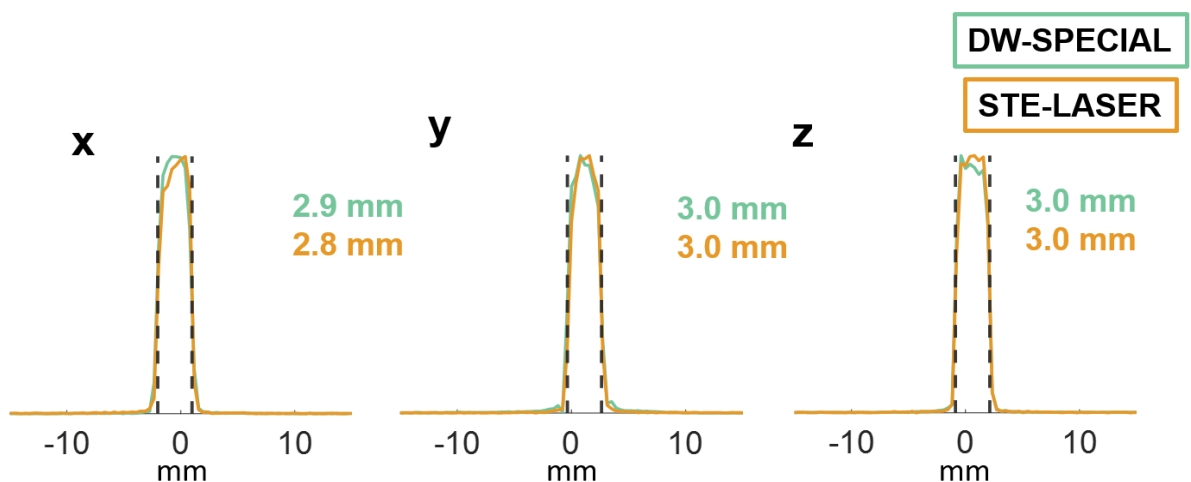

*Figure S5 – 1D projections of the x, y, z selection profiles for a small voxel (3x3x3 mm³) in a $B_1$-homogeneous region measured with DW-SPECIAL (green) and STE-LASER (orange) in the multi-metabolite phantom. The profiles are obtained by switching on a gradient during the acquisition. The integral values of the profile shapes are displayed. The dashed black lines represent the nominal voxel position in each direction. Negligible difference is observed in the voxel selection between the two sequences when the factor of $B_1$ inhomogeneity is removed, confirming the good selection performed with DW-SPECIAL.*

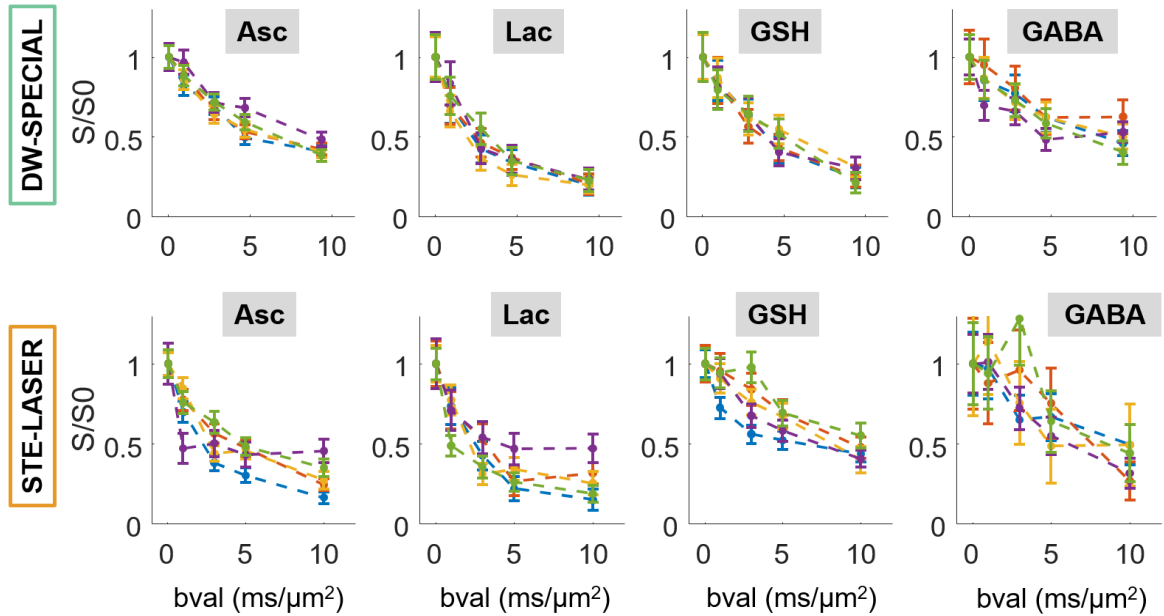

*Figure S6* – Concentration decays of Asc, Lac, GSH and GABA as a function of b-value for all animals (different colors) and both sequences, normalized to their value at b = 0.05 ms/μm². Error bars: absolute CRLB from LCModel quantification. Although these metabolites are generally not reported in DWS studies owing to their low concentration and poor quantification, the shorter TE achieved in DW-SPECIAL leads to a smaller within-group dispersion of their diffusion decays and suggests that these metabolites could possibly be investigated in future studies (through a fit of the mean diffusion decay).

| DW-SPECIAL – LCModel concentrations (arbitrary units) | | | | | | DW-SPECIAL – LCModel relative CRLB | | | | | |
|---|---|---|---|---|---|---|---|---|---|---|---|
| Metabolite | b = 0.05 ms/μm² | b = 1 ms/μm² | b = 3 ms/μm² | b = 5 ms/μm² | b = 10 ms/μm² | Metabolite | b = 0.05 ms/μm² | b = 1 ms/μm² | b = 3 ms/μm² | b = 5 ms/μm² | b = 10 ms/μm² |
| Gln | 4,8 | 4,2 | 3,1 | 2,5 | 1,5 | Gln | 0,064 | 0,066 | 0,078 | 0,09 | 0,116 |
| Glu | 16,4 | 14 | 10,8 | 8,7 | 6,8 | Glu | 0,02 | 0,02 | 0,022 | 0,026 | 0,028 |
| mIns | 12,8 | 11,1 | 8,8 | 7 | 5,2 | mIns | 0,02 | 0,02 | 0,02 | 0,024 | 0,03 |
| Tau | 9,9 | 8,1 | 5,9 | 4,4 | 3,2 | Tau | 0,024 | 0,028 | 0,03 | 0,036 | 0,042 |
| NAA+NAAG | 14,1 | 12,2 | 9,5 | 7,7 | 5,7 | NAA+NAAG | 0,014 | 0,018 | 0,018 | 0,02 | 0,02 |
| Cr+PCr | 14,2 | 12 | 9 | 7 | 4,8 | Cr+PCr | 0,01 | 0,012 | 0,016 | 0,022 | 0,034 |
| GPC+PCho | 1,7 | 1,4 | 1,1 | 0,9 | 0,6 | GPC+PCho | 0,04 | 0,044 | 0,046 | 0,056 | 0,068 |
| MM | 2,3 | 2,3 | 2,2 | 2,1 | 2,1 | MM | 0,022 | 0,022 | 0,02 | 0,02 | 0,018 |
| Asc | 7 | 6,2 | 4,8 | 4 | 2,9 | Asc | 0,052 | 0,054 | 0,06 | 0,066 | 0,072 |
| Lac | 1,6 | 1,2 | 0,7 | 0,5 | 0,3 | Lac | 0,1 | 0,124 | 0,176 | 0,216 | 0,284 |
| GSH | 1,4 | 1,2 | 0,9 | 0,7 | 0,4 | GSH | 0,106 | 0,114 | 0,13 | 0,156 | 0,234 |
| GABA | 2,4 | 2 | 1,8 | 1,4 | 1,2 | GABA | 0,1 | 0,11 | 0,11 | 0,122 | 0,124 |

*Table S1* - Concentrations and relative Cramer Rao Lower Bounds from LCModel fit of DW-SPECIAL spectra averaged over animals at all b-values and for every reported metabolite. Metabolite concentrations are not referenced to water and their concentrations are in arbitrary unit: only their relative contributions at b = 0.05 ms/μm² is of biological relevance. The MM basis-set spectrum was scaled such that its concentration at b = 0.05 ms/μm² is in the range 1-4mM (Cudalbu et al., https://doi.org/10.1002/nbm.4393).

| STE-LASER – LCModel concentrations (arbitrary units) | | | | | | STE-LASER – LCModel relative CRLB | | | | | |
|---|---|---|---|---|---|---|---|---|---|---|---|
| Metabolite | b = 0.05 ms/µm² | b = 1 ms/µm² | b = 3 ms/µm² | b = 5 ms/µm² | b = 10 ms/µm² | Metabolite | b = 0.05 ms/µm² | b = 1 ms/µm² | b = 3 ms/µm² | b = 5 ms/µm² | b = 10 ms/µm² |
| Gln | 4 | 3,3 | 2,4 | 2 | 1,4 | Gln | 0,082 | 0,09 | 0,112 | 0,134 | 0,166 |
| Glu | 14,8 | 12,1 | 9 | 7,5 | 5,3 | Glu | 0,02 | 0,02 | 0,026 | 0,032 | 0,04 |
| mIns | 10,8 | 9,1 | 7,2 | 6 | 4,3 | mIns | 0,02 | 0,02 | 0,02 | 0,024 | 0,032 |
| Tau | 8 | 6,4 | 4,6 | 3,6 | 2,4 | Tau | 0,02 | 0,024 | 0,03 | 0,036 | 0,05 |
| NAA+NAAG | 13,5 | 11,3 | 8,7 | 7,4 | 5,5 | NAA+NAAG | 0,01 | 0,01 | 0,016 | 0,018 | 0,022 |
| Cr+PCr | 12 | 9,9 | 7,6 | 6,1 | 4,2 | Cr+PCr | 0,01 | 0,01 | 0,01 | 0,018 | 0,022 |
| GPC+Pcho | 2,2 | 1,9 | 1,5 | 1,2 | 0,8 | GPC+PCho | 0,028 | 0,032 | 0,026 | 0,04 | 0,054 |
| MM | 3 | 2,9 | 2,8 | 2,8 | 2,7 | MM | 0,03 | 0,028 | 0,028 | 0,024 | 0,022 |
| Asc | 4,6 | 3,3 | 2,3 | 2 | 1,3 | Asc | 0,062 | 0,09 | 0,106 | 0,118 | 0,164 |
| Lac | 1,9 | 1,3 | 0,8 | 0,6 | 0,5 | Lac | 0,092 | 0,118 | 0,174 | 0,246 | 0,268 |
| GSH | 1,9 | 1,6 | 1,4 | 1,2 | 0,8 | GSH | 0,07 | 0,072 | 0,08 | 0,092 | 0,124 |
| GABA | 2 | 2 | 1,7 | 1,3 | 0,8 | GABA | 0,176 | 0,162 | 0,176 | 0,242 | 0,342 |

*Table S2 - Concentrations and relative Cramer Rao Lower Bounds from LCModel fit of STE-LASER spectra averaged over animals at all b-values and for every reported metabolite. Metabolite concentrations are not referenced to water and their concentrations are in arbitrary unit: only their relative contributions at b = 0.05 ms/µm² is of biological relevance. The MM basis-set spectrum was scaled such that its concentration at b = 0.05 ms/µm² is in the range 1-4mM (Cudalbu et al., https://doi.org/10.1002/nbm.4393).*

| 1. Hardware | |
|---|---|
| a. Field strength [T] | 14 T |
| b. Manufacturer | Bruker |
| c. Model (software version if available) | Avance Neo, Paravision 360 v1.1 |
| d. RF coils: nuclei (transmit/receive), number of channels, type, body part | Homemade quadrature 2 loops surface coil (2 cm diameter for each loop) |
| e. Additional hardware | Gradient strength: 1T/m, rise time: 270µs |
| 2. Acquisition | |
| a. Pulse sequence | DW-SPECIAL and STE-LASER |
| b. Volume of Interest (VOI) locations | Full brain (voxel location on Fig 2. A) |
| c. Nominal VOI size [cm³, mm³] | 7x5x5mm³ |
| d. Repetition Time (TR), Echo Time (TE), mixing time (TM) | DW-SPECIAL:<br>TE: 18.5 ms<br>TM: 40 ms<br>TR: 3000 ms |

|  | STE-LASER:<br>TE: 33 ms<br>TM: 40 ms<br>TR: 3000 ms |
|---|---|
| e. Total number of Excitations or acquisitions per spectrum | 160 shots per b-value for b<10 ms/µm$^2$<br>320 shots for b = 10 ms/µm$^2$<br>5 b-values from 0.05 to 10 ms/µm$^2$ |
| f. Additional sequence parameters (spectral width in Hz, number of spectral points, frequency offsets) | 7142 Hz<br><br>4096 points<br><br>Duration of diffusion gradients: 3ms |
| g. Water Suppression Method | VAPOR |
| h. Shimming Method, reference peak, and thresholds for "acceptance of shim" chosen | MAPSHIM and local iterative shimming in the MRS voxel, target LW: 17-19Hz |
| i. Triggering or motion correction method | None |
| **3. Data analysis methods and outputs** | |
| a. Analysis software | LCModel v6.3 |
| b. Processing steps deviating from quoted reference or product | EC, phase and frequency drifts correction (spectral registration in FID-A), motion-corrupted shots removal |
| c. Output measure (e.g. absolute concentration, institutional units, ratio) | Concentrations in arbitrary units |
| d. Quantification references and assumptions, fitting model assumptions | The basis set includes an in vivo acquired MM spectrum<br>NUNFIL 2048<br>NRATIO 0<br>NSIMUL 0<br>Fit region: 0.2-4.3ppm |
| **4. Data Quality** | |
| a. Reported variables (SNR, Linewidth (with reference peaks)) | SNR: not reported<br>LW: not evaluated<br>Mean and SD across individual animal diffusion decays, CRLB and estimated diffusion parameters.<br>Agreement between these diffusion parameters and the ones derived from the fit of the mean decay. |
| b. Data exclusion criteria | Shots whose MSE with the median of the shots in frequency domain deviating from more than 1.5 SD from the mean MSE are discarded |
| c. Quality measures of postprocessing Model fitting (e.g. CRLB, goodness of fit, SD of residual) | SD across animals and LCModel CRLB |
| d. Sample Spectrum | Figure 2 and 3 |

*Table S3* - *MRSinMRS checklist from Lin et al. « Minimum Reporting Standards for in Vivo Magnetic Resonance Spectroscopy (MRSinMRS): Experts' Consensus Recommendations ».*